\documentclass[letterpaper,10pt,twocolumn]{article}
\pdfoutput=1
\usepackage{geometry} 
\usepackage{times}
\usepackage{datetime}
\usepackage{amsmath}
\usepackage{epsfig,endnotes,url,xspace}
\usepackage{enumitem,comment,graphicx,subfigure}
\usepackage[small]{caption}
\usepackage{multirow}
\usepackage{algorithm}
\usepackage{algorithmic}
\usepackage{amsthm}

\newcommand{\ie}{{\em i.e.,}~}
\newcommand{\eg}{{\em e.g.,}~}
\newcommand{\smallurl}[1]{{\url{#1}}}
\newcommand{\eat}[1]{}

\title{\bf Faster and More Accurate Sequence Alignment with SNAP}

\author{Matei Zaharia, William J.~Bolosky$^\dag$, Kristal Curtis, Armando Fox, David Patterson,\\
Scott Shenker, Ion Stoica, Richard M.~Karp, Taylor Sittler$^\ddag$\footnote{\emph{Corresponding Author:} Department of Laboratory Medicine, University of California San Francisco 
185 Berry Street, Suite 100 
San Francisco, CA 94115. 
\texttt{taylor.sittler@ucsf.edu}}\\
\\
U.C.~Berkeley $\;\;\;$ $^\dag$Microsoft Research $\;\;\;$ $^\ddag$U.C.~San Francisco\\
}

\date{}

\begin{document}

\maketitle


\abstract{
\small
We present the Scalable Nucleotide Alignment Program (SNAP), a new short and long read aligner
that is both more accurate (\ie aligns more reads with fewer errors) and
10--100$\times$ faster than state-of-the-art tools such as BWA.
Unlike recent aligners based on the Burrows-Wheeler transform, SNAP uses a simple hash
index of short seed sequences from the genome, similar to BLAST's.
However, SNAP greatly reduces the number and cost of local alignment checks performed
through several measures: it uses longer seeds to reduce the false positive locations considered,
leverages larger memory capacities to speed index lookup, and excludes most candidate locations
without fully computing their edit distance to the read. 
The result is an algorithm that scales well for reads from one hundred to thousands of bases
long and provides a rich error model that can match classes of mutations (\eg longer
indels) that today's fast aligners ignore.
We calculate that SNAP can align a dataset with 30$\times$ coverage of a
human genome in less than an hour for a cost of \$2 on Amazon EC2, with
higher accuracy than BWA.
Finally, we describe ongoing work to further improve SNAP.
}


\section{Introduction}

As massively parallel sequencing methods become common in biological research, high throughput sequence information is making its way into a variety of fields, from plant biology to human infectious disease, cancer research, and clinical medicine.  With the advent of newer sequencing machines, hundreds of millions to billions of short nucleotide fragments are now generated in a single experiment.  

Given recent trends and the current demand for these technologies, monitoring the cost of sequencing provides a good measure of the amount of sequence generated (lower cost means more sequencing).  That cost is decreasing vanishingly fast.  Resequencing a human genome cost several billion dollars at the turn of the millennium, reduced to \$10,000 at the start of 2011 \cite{seqcosts}.  By year's end, the equivalent genome is projected to cost only \$2,000.  Far outstripping Moore's Law, the amount of generated sequence threatens to overwhelm current storage and compute infrastructures.  Laboratories and sequencing centers are turning to ``the cloud":  renting machines and storage from companies like Amazon, Rackspace, and Microsoft.  However, given the rate of decrease in cost and increase in the amount of sequence data, simply scaling out across more machines cannot keep pace with these advances.  New algorithms are required to handle this data.

The most popular traditional algorithms for sequence alignment are Smith-Waterman and BLAST~\cite{blast}.  By themselves, these algorithms are quite accurate, but they take an inordinate amount of time to perform adequate alignments.  As a result, many new aligners have been developed for the purpose of handling large amounts of short read data over the past two to four years.  BWA, SOAP, Bowtie and SSAHA~\cite{bwa,soap2,bowtie,ssaha} are but a few recent examples of such tools.  

While these aligners outperform Smith-Waterman and BLAST by several orders of magnitude, they are still computationally expensive, taking multiple CPU-days to align a single human genome.  Furthermore, the faster aligners' speed comes at the cost of accuracy: these aligners only support limited numbers and types of errors, \eg one insertion/deletion (indel) of limited size per read.  They can thus miss larger differences between sequences that are biologically significant.
For example, short insertions and deletions are thought to comprise between 15-30\% of known genetic variation~\cite{mills2006initial} and to contribute to common diseases~\cite{vallania2010high}.  Additionally, human cancer cells often contain a host of anomalies~\cite{balmain2003genetics}, including important base substitutions and short indels.  As DNA repair mechanisms degrade in tumor cells, accumulation of these mutations is thought to drive cancer progression.  Accurate identification of these changes is therefore critical to understanding these diseases.

In this article, we present SNAP, a new aligner that is both substantially faster and more accurate than the current state-of-the-art algorithms.  SNAP has several properties that make it attractive and broadly applicable for truly high throughput sequence analysis:
\begin{enumerate}
    \item It runs 10--100$\times$ faster than existing tools on reads from current sequencing technologies, while providing higher accuracy (more reads aligned with fewer errors).
    \item It supports a rich error model: SNAP can find alignments with an arbitrary number of substitutions, insertions, or deletions from the reference genome, as long as there is one contiguous ``seed'' of $\sim$20 bases matching exactly.
    \item It can run on readily available hardware, such as several Amazon EC2 server types.
    \item The same algorithm can be used across a wide range of read lengths (from 100 to 10000 base pairs) and error rates, making it applicable to both current an upcoming sequencing technologies.
\end{enumerate}

Surprisingly, unlike the fastest current aligners (\eg Bowtie and BWA), which search a trie index of the genome encoded using the Burrows-Wheeler transform, SNAP uses a simple hash index similar to BLAST's, based on short ``seed'' substrings of the genome.
However, SNAP leverages several observations to run faster than previous hash-based aligners. 
First, read lengths have increased since the advent of BLAST: whereas sequencers used to produce 25--30 base pair reads, they are now producing reads of 100 bp or more. This lets us use longer seeds ($\sim$20 bases as opposed to BLAST's 10--12), which have a higher probability of containing an error and thus not being found in the index, but match far fewer ``false positive" locations by chance. 
Second, we eliminate most of these locations without fully computing their edit distance from the read, by
using a local alignment algorithm that can quickly reject locations with a higher edit distance than the
best we found so far~\cite{ukkonen} and eliminating some locations based on number of matching seeds.
This gives us a 10--50$\times$ speedup over the textbook $O(n^2)$ edit distance check~\cite{edit-distance-alg}.
Third, SNAP leverages the higher memory capacities on today's servers to index more seeds and perform fewer hash lookups.\footnote{The current version uses 39 GB for an index of the human genome, which is readily available on commodity servers and on cloud services. We believe that the memory usage could be reduced with little loss in speed, as most of the algorithm's time is spent in local alignment rather than index lookup.}

Combined, these optimizations yield an algorithm that would not have performed well on the ultrashort reads available when BLAST was introduced, but is highly effective on the 100+ bp reads available today, as well as next-generation 1000+ bp long read technology (\eg Pacific Biosciences).
Indeed, we were able to align a 100 bp read dataset with 30-fold coverage of a human genome in 20 minutes
on a 32-core server that cost \$11,000 in November 2011, using parameters that provide higher accuracy
than BWA, and we estimate that the same dataset could be aligned in less than an hour for \$2.40 on
a ``cc2.8xlarge'' Amazon EC2 server.
More importantly, since trends in both current and next-generation sequencing technologies indicate that read lengths will keep increasing, we believe that the techniques in SNAP will continue to be relevant in the future.


\section{Results}

Before presenting the SNAP algorithm in detail, we start with a speed and accuracy
comparison against existing aligners.
We evaluated the algorithms using simulated reads, for which we can know the true genome location
and can compute error rates. We computed three metrics:
\begin{itemize}
  \item \emph{Aligned reads:} Percent of reads that the aligner confidently mapped to a location.
  \item \emph{Errors:} Percent of confidently aligned reads that are 
  mapped to the wrong location.
  \item \emph{Speed:} Reads per second aligned on a single CPU core. All the aligners
  parallelize on machines with multiple cores.
\end{itemize}

To determine confident reads, we used a quality threshold of 10 for the aligners
that report quality scores. This value was also used in the evaluation of BWA~\cite{bwa}.

We start by showing results for short simulated reads (100 and 200 bp) with mutation
frequencies matching human data (Section \ref{sec:results-short}).
These reads are representative of current sequencing technologies,
such as Illumina.
Next, we compare performance on simulated long read data with high indel rates from
sequencing errors, consistent with ``third-generation" sequencing technologies
 (Section \ref{sec:results-long}).
Finally, we report preliminary results for a multicore version of SNAP
(Section \ref{sec:results-parallel}).
Unless otherwise noted, the measurements are from an ``m2.4xlarge'' Amazon EC2 machine with 68 GB RAM.

\subsection{Short Reads}
\label{sec:results-short}

\begin{table*}[t]
  \centering
  \small

  \begin{tabular}{|c|c|r|r|r|r|r|r|}
    \hline
    \multirow{2}{*}{Error rate} & \multirow{2}{*}{Program} &
    \multicolumn{3}{|c|}{100 bp} & \multicolumn{3}{|c|}{200 bp} \\ \cline{3-8}
     & & \% Aligned & \% Error & Reads/s & \% Aligned & \% Error & Reads/s \\ \hline
    \multirow{4}{*}{2\%}
      & BWA    & 90.8 & 0.04 &    942 & 91.7 & 0.02 &    430 \\
      & SOAP2  & 93.7 & 1.53 &  1,920 & 87.9 & 1.14 &    570 \\
      & Bowtie & 88.7 & 1.08 &    368 & 91.8 & 0.40 &    935 \\
      & SNAP   & 92.0 & 0.05 & 28,400 & 94.4 & 0.03 & 35,800 \\ \hline
    \multirow{4}{*}{5\%}
      & BWA    & 52.9 & 0.16 &    782 & 27.8 & 0.06 &    702 \\
      & SOAP2  & 73.4 & 1.92 &    665 & 18.2 & 1.15 &    215 \\
      & Bowtie & 76.4 & 2.18 &    241 & 85.7 & 0.78 &    624 \\
      & SNAP   & 87.4 & 0.09 & 11,300 & 92.3 & 0.04 & 13,700 \\ \hline
    \multirow{4}{*}{10\%}
      & BWA    &  4.6 & 0.42 & 2,000 &  0.1 & 0.10 & 6,250 \\
      & SOAP2  & 17.9 & 2.16 &   665 &  0.1 & 1.19 &   423 \\
      & Bowtie & 40.5 & 5.67 &   161 & 53.7 & 2.18 &   416 \\
      & SNAP   & 70.7 & 0.48 & 4,600 & 82.7 & 0.14 & 5,170 \\ \hline  
  \end{tabular}
  
  \caption{Results from aligning short reads with various tools.
  One million simulated reads were generated from the human genome using a 0.09\% SNP 
  mutation rate, 0.01\% indel mutation rate, and varying sequencing error rates.
  The metrics show fraction of reads aligned, fraction of misalignments out of the
  aligned reads, and speed on a single CPU, averaged over three runs.}
  \label{tbl:results-short}
\end{table*}

We compared SNAP's performance on short reads against Bowtie~\cite{bowtie}, SOAP2~\cite{soap2}
and BWA~\cite{bwa}. We created one million simulated reads of 100 and 200 bp from the
\texttt{hg19} human reference genome, using the \texttt{wgsim} program in SAMtools~\cite{samtools}.
We simulated a 0.1\% mutation rate (0.09\% SNPs and 0.01\% indels) representative of human data,
and base sequencing error rates of 2\%, 5\% and 10\%, similar to the evaluation in~\cite{bwa}.

Table~\ref{tbl:results-short} reports our results. We see that the existing aligners vary
in terms of speed and accuracy, with SOAP2 being the fastest but the least accurate,
and BWA being slightly slower but more accurate. However, SNAP performs 10--50$\times$
faster than these tools while also aligning more reads and making fewer errors.
Furthermore, SNAP continues to perform well when the error rate increases, whereas the
number of reads aligned by BWA, SOAP2 and Bowtie drops substantially because these algorithms
are designed for small numbers of errors per read.
In practice, this means that SNAP can also match reads with more \emph{mutations} collected
using lower-error sequencing technologies.

For a direct comparison with other reports of aligner performance, we also ran a test with 125 bp reads
simulated as in the BWA paper~\cite{bwa}. The paper reports that BWA aligned 93.0\% of the reads
at 0.05\% error and 662 reads/s, while we found that SNAP aligned 94.1\% of the reads at 0.05\% error
and 34100 reads/s.

\subsection{Long Reads}
\label{sec:results-long}

Third-generation sequencing technologies, such as Ion Torrent or Pacific Biosciences, will generate
long reads of several thousand base pairs with a high frequency of indels due to sequencing errors.
Intuitively, SNAP should work well for these types of reads: a larger read length allows us to safely use long
seeds in hashtable lookups (see Section~\ref{sec:methods-index}) and find good candidate locations
quickly, while the change to indels from substitutions does not affect our edit distance algorithm.

To evaluate SNAP for this type of data, we generated reads with varying sequencing error rates,
in which 20\% of the sequencing errors were indels and the other 80\% were substitutions, as in
the evaluation of BWA-SW~\cite{bwa-sw}. We then compared SNAP against BWA-SW.

\begin{table*}[t]
  \centering
  \small
  
  \begin{tabular}{|c|c|r|r|r|r|r|r|}
    \hline
    \multirow{2}{*}{Error rate} & \multirow{2}{*}{Program} &
    \multicolumn{3}{|c|}{1000 bp} & \multicolumn{3}{|c|}{10,000 bp} \\ \cline{3-8}
     & & \% Aligned & \% Error & Reads/s & \% Aligned & \% Error & Reads/s \\ \hline
    \multirow{2}{*}{2\%}
      & BWA-SW & 96.2 & 0.02 &   49 & 97.1 & 0.00 &   7 \\
      & SNAP   & 96.9 & 0.02 & 8020 & 98.3 & 0.01 & 313 \\ \hline
    \multirow{2}{*}{5\%}
      & BWA-SW & 96.0 & 0.04 &   58 & 97.1 & 0.01 &   8 \\
      & SNAP   & 96.6 & 0.04 & 2180 & 97.9 & 0.01 & 116 \\ \hline
    \multirow{2}{*}{10\%}
      & BWA-SW & 95.8 & 0.03 &   81 & 97.5 & 0.19 &   9 \\
      & SNAP   & 95.9 & 0.03 &  660 & 97.7 & 0.04 &  33 \\ \hline
  \end{tabular}
  
  \caption{Results for long simulated reads with varying sequencing error rates.
  20\% of the sequencing errors were indels.}
  \label{tbl:results-long}
\end{table*}

Table~\ref{tbl:results-long} shows the results. We see that SNAP performs substantially
faster than existing tools---from as much as 164$\times$ faster for the 1000 bp reads
with 2\% error, to 3.6$\times$ faster for 10,000 bp with 10\% error. We have not yet optimized
SNAP's local alignment performance for very long reads, so we believe
that we can further improve performance in these cases. Crucially, however, SNAP's accuracy
also exceeds that of existing tools: it often aligns 0.5\% to 1\% more reads, while achieving
a similar or smaller error rate.

\subsection{Multi-Core Implementation}
\label{sec:results-parallel}

We built a parallel version of SNAP that runs alignments on all of the cores of a multi-processor.  It works by assigning a range of the input file to a core, having the core process the entire range and then taking another.  The size of each the ranges is about half of the total remaining work divided by the number of cores, with a minimum size that is about a half second's work.  This scheme balances the efficiency of having large chunks (and so amortizing the per-chunk overhead over more work) with the desire to have all of the cores finish at the same time, without having some exit early for lack of work while others are still processing.  Our preliminary implementation works well even though we've observed some cores running at $1.7\times$ the speed of some others.

Because alignments of different reads do not depend on one another, one might think that running an alignment on multiple cores of a multiprocessor machine would result in perfect speedup.  However, while the individual alignments are logically independent, the computational resources used to process them are not.  The cores contend for memory bandwidth to access the index, reference genome and input file; the cores on a single chip contend for L3 cache; and, even though there is almost no shared memory that is written, the memory system still has to run a coherence protocol to verify that there have been no writes.  All of these effects reduce the scaling.   We have not yet tried to quantify their relative effects, nor have we tuned the parallel version of SNAP.

We ran tests on a Dell machine with four 8-core 2.6~GHz AMD Opteron 6140 processors, for a total of 32 cores.  The machine has 256GB of memory and an IO system consisting of 6 1TB 2.5'' 7200 RPM SAS disks attached to a PERC H700 RAID controller configured as one drive for the operating system and a 5 disk RAID-0 stripe to hold the indices and read data.  This machine cost about \$11,000 in November of 2011.

\begin{figure}
\centering
\includegraphics[width=3.05in]{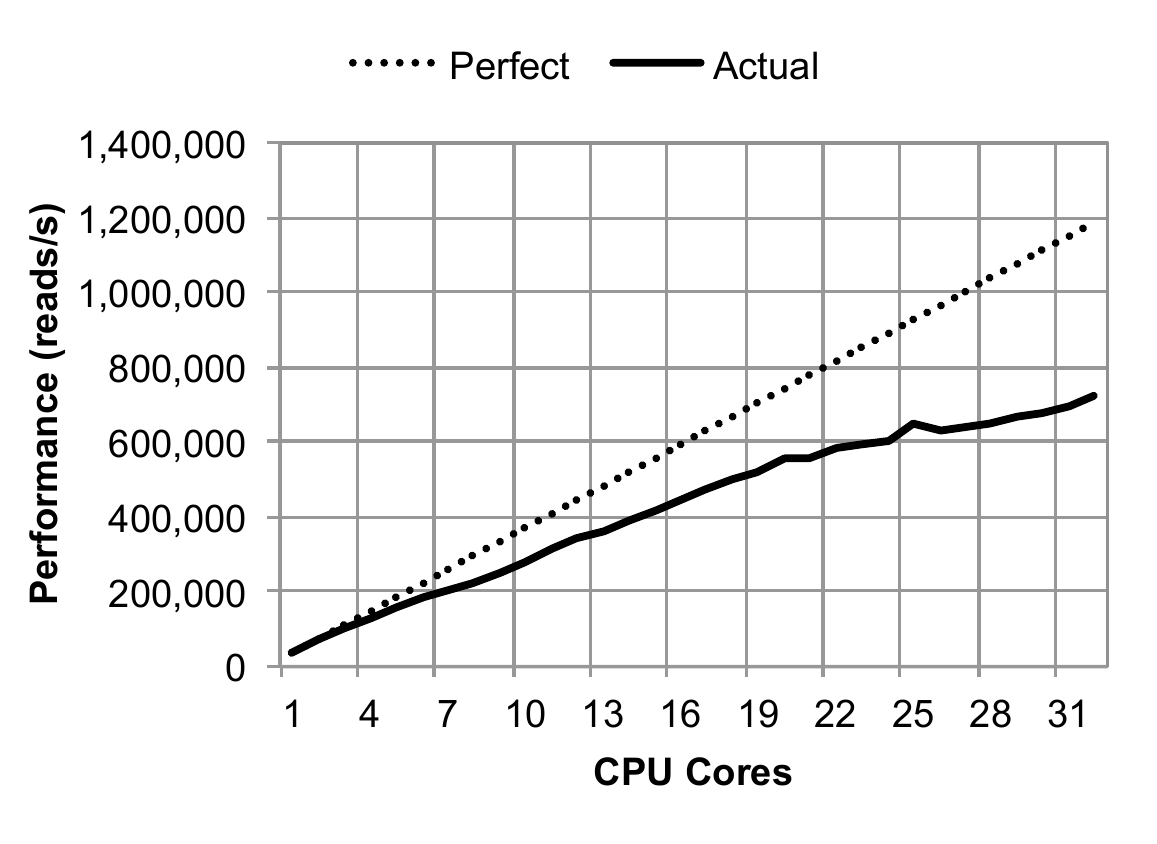}
\caption{Scaling performance of SNAP in aligning one hundred million 125-base
simulated reads on multiple cores.}
\label{fig:multicore-scalability}
\end{figure}

To measure scaling, we ran a simulated dataset containing 100 million 125-base reads with a 2\% error rate, while varying the number of cores used.  SNAP produced about 94\% confident alignments with an error rate of 0.05\%.  On a single core, SNAP ran at about 37,000 reads/s. On all 32 cores, it ran at 723,000 reads/s, just under a factor of 20 faster, or about 60\% of perfect speedup.  Figure \ref{fig:multicore-scalability} shows the scaling performance.

We also analyzed NA12878 (the mother) from the CEPH Trio dataset~\cite{ceph-trio}, treating the data as single-end
reads.  In total, there were 1.1 billion such reads, each 100 bp long.
SNAP processed them at a rate of approximately 1M reads/s on 32 cores, and got through the entire set of reads in under 20 minutes with a confident match rate of about 79\%.  
At this speed, we calculate that the 16-core, 60~GB RAM ``cc2.8xlarge" server type available on Amazon EC2, which costs \$2.40 per hour, can align the dataset in about 40 minutes.


\section{Methods}
\label{sec:methods}

Like the original BLAST algorithm~\cite{blast}, SNAP is based on a hash index of short substrings
of the genome (or other reference database) called \emph{seeds}, of a fixed size $s$. Given a read to align, SNAP draws multiple substrings of length $s$ from it and performs an exact lookup in the hash index to
find locations in the database that contain the same substrings. It then computes the edit
distance between the read and each of these \emph{candidate locations} to find the best alignment.
Two features differentiate SNAP from previous algorithms, however, and give it its speed:
the use of a larger index that trades memory to reduce computation, and a set of
optimizations for the local alignment step that greatly reduce the cost of testing a
read against its candidate locations.

\subsection{Hash Index}
\label{sec:methods-index}

SNAP's index is a hash table from seed strings of a fixed length $s$ to lists of positions in the
reference database where those strings occur. It differs from the indices in BLAST in two ways:

\paragraph{1. Longer Seeds:}
We chose to index longer seeds than most existing tools, with typical values of 
$s$ being 20 base pairs in contrast to BLAST's 10--12 bp. Longer seeds mean that it is less
likely for a seed we extract from a read to exactly match the reference database (\eg with a
sequencing error rate of 2\%, the probability that a 20 bp seed is error-free is
$0.98^{20}=0.67$), but also drastically reduces the number of \emph{false positive} seed
matches. For example, given that the human genome is approximately $4^{16}$ bp, we
expect a seed of length 10, as in BLAST, to match $4^6=4096$ locations just by chance,
resulting in many unneeded local alignment tests. In contrast, with a seed of length 20,
the expected number of hits by chance is $4^{-4}=0.0039$. Previous tools avoided long
seeds mostly due to reads being shorter (\eg for a 35 bp read with an error in the
middle, all 20 bp seeds would contain the error), but with 100 bp and larger reads, it
is possible to draw enough independent long seeds to have a high chance of one being error-free.
For example, from a 100 bp read, we can draw five non-overlapping 20 bp seeds, and with
a 2\% error rate, the chance that any one seed has an error is $1-0.67=0.33$; however, the chance
that \emph{all} five seeds have errors is only $0.33^5 = 0.004$.

\paragraph{2. Overlapping Seeds:} To save memory, many hash-based aligners index only non-overlapping $s$-mers of the reference database (\eg the first 10 bases, then the next 10
bases, etc), so that with a seed length of $s$, only $1/s$ of the locations in the database
occur in the index. For example, an index over the human genome with 20 bp seeds would
take about 2 GB of memory with non-overlapping seeds. However, the disadvantage of this
approach is that one must try more seeds per read to find one that matches the stride
used in the index (\eg only one in 20 seeds in the read can match the
seeds used in the index). While the computational cost of additional hash table lookups seems
like it ought to be small, it is in fact substantial, because each lookup will result in
a CPU cache miss due to the index being much larger than the processor's L3 cache.\footnote{
L3 caches in modern processors are only several MB in size.} A read from main memory, without
going through the processor cache, takes hundreds of cycles in modern processors.
SNAP avoids this cost by indexing \emph{overlapping} $s$-mers (\ie sliding a window of length
$s$ one base at a time over the database). This requires a larger index, but one
that is still well within the means of modern servers; for example, SNAP requires
39 GB for an index of the human genome.\footnote{Note that it would also be possible to choose 
strategies in-between SNAP's fully overlapping windows and traditional disjoint windows.
For example, we could slide a window 2 bases at a time, resulting in a 2$\times$ smaller index
but more hash table lookups. We chose to to index all
windows because the memory required was manageable.}

\subsection{Local Alignment}

Although our use of longer seeds and a large index substantially reduces the number of local
alignment checks SNAP performs, we still found that the algorithm spent most of its cycles
in local alignment---that is, in computing the edit distance from the read to the
candidate locations where we found matching seeds. SNAP goes further to reduce the number
and cost of local alignment checks by leveraging three observations:

\begin{enumerate}
  \item With the longer seeds used in our index, there will be few false positive hits,
  so we will find a good alignment for each read quickly.
  
  \item For most applications, one only needs to find the best and second-best alignments
  of a read, \ie the positions where the edit distance between the read and the
  database is smallest and second-smallest. If the best alignment is sufficiently better than
  the second-best, then the read unambiguously maps to that location, but if both 
  alignments are good, the read cannot be mapped confidently to either place.
  
  \item We can eliminate some locations solely on the number of seeds that match there,
  without performing an edit distance check. For example, if three of the non-overlapping
  seeds from a read do not match with a particular candidate location, then that candidate
  location must have at least edit distance 3 from the read, because there must be at
  least one difference in each of the non-matching seeds. This observation was first made
  by Baeza-Yates~\cite{baeza-yates-seeds}.
\end{enumerate}

Algorithm~\ref{alg:aligner} shows the pseudocode for SNAP, which uses all three
observations to eliminate candidates. The algorithm takes five parameters controlling
which alignments it considers good and which seeds it tries, shown in Table~\ref{tbl:parameters}.
It return one of three values for each read: ``single hit,''
if it found only one good alignment for the read; ``multiple hits,'' if it finds
two or more alignments that are too close in score; and ``not found.''\footnote{
We differentiate between ``multiple hits" and ``not found" to let SNAP be used for filtering in
addition to precise alignment, \eg to filter out human reads from data collected in search
for a pathogen even if these reads match more than one location in the human genome.}


To leverage our first two observations, SNAP uses an edit distance algorithm that runs
faster when comparing more similar strings, and can return without fully computing
the distance if it exceeds a given threshold.
The textbook dynamic programming algorithm for
edit distance~\cite{edit-distance-alg} takes $O(n^2)$ time to compare two
strings of length $n$. Instead, we
use an algorithm by Ukkonen~\cite{ukkonen} whose running time is $O(nd)$, where $d$
is the edit distance between the strings, and whose space cost is $O(d)$.
Furthermore, if $d$ exceeds a given threshold,
$d_\text{limit}$, the algorithm returns a special value in $O(nd_\text{limit})$ time
using $O(d_\text{limit})$ space, without fully computing the distance.\footnote{
The algorithm is roughly equivalent to filling in only the central diagonals of the
traditional $n^2$ size edit distance table, but it only tracks the farthest
distance one can travel on each diagonal to save space.
}

\begin{table}[t]
  \centering
  \small
  \begin{tabular}{|p{0.8in}|p{2.05in}|}
    \hline
    \textbf{Parameter} & \textbf{Meaning} \\
    \hline
    Seed size ($s$) & Length of seeds, in bases. \\ \hline
    Seeds to try ($n$) & Number of seeds to try for each read. \\ \hline
    Maximum distance ($d_\text{max}$) & Maximum edit distance from reference
                                    sequence to allow for an alignment. \\ \hline
    Confidence threshold ($c$) & Difference in edit distances between
                                 a read's best and second-best alignments
                                 needed to report it as confidently aligned. \\ \hline
    Max hits ($h_\text{max}$) & Maximum index locations to check for a seed.
                                Some seeds containing repetitive strings
                                (\eg \texttt{AAAAAA}) hit thousands of locations,
                                so we ignore them. \\ \hline
  \end{tabular}
  \caption{SNAP algorithm parameters.}
  \label{tbl:parameters}
\end{table}

\begin{algorithm}[t]
\caption{SNAP alignment algorithm.}
\label{alg:aligner}
\small
\begin{algorithmic}
\STATE $d_\text{best} \leftarrow \infty$
\STATE $d_\text{second} \leftarrow \infty$
\FOR{$i = 1 \dots n$}
  \STATE $S \leftarrow i^\text{th}$ seed of read
  \IF{\# of index entries for $S \le h_\text{max}$}
    \FOR{$l \in $ locations of $S$ in index}
      \STATE $p \leftarrow l - \text{offset of seed $i$ from start of read}$
      \STATE $\text{SeedsHitting[$p$]} \leftarrow \text{SeedsHitting[$p$]} + 1$
    \ENDFOR
    \STATE $p \leftarrow$ unscored location with the most seeds hitting
    \IF {$d_\text{best} > d_\text{max}$}
      \STATE $d_\text{limit} \leftarrow d_\text{max} + c - 1$
    \ELSIF {$d_\text{second} \ge d_\text{best} + c$}
      \STATE $d_\text{limit} \leftarrow d_\text{best} + c - 1$
    \ELSE
      \STATE $d_\text{limit} \leftarrow d_\text{best} - 1$
    \ENDIF
    \STATE $d \leftarrow \text{EditDistance}(\text{Read}, \text{Reference}[p], d_\text{limit})$
    \STATE update $d_\text{best}$ and $d_\text{second}$ based on newly scored $d$
    \IF{$d_\text{best} < c$ and $d_\text{second} < d_\text{best} + c$}
      \STATE {\bf return} multiple hits (we have two hits within distance $c$ and no better hit can be confident)
    \ELSIF{\# non-overlapping seeds tested $\ge d_\text{best} + c$}
      \STATE score remaining locations and break (any unscored location will have too high a distance)
    \ENDIF
  \ENDIF
\ENDFOR
\IF{$d_\text{best} \le d_\text{max}$ and $d_\text{second} \ge d_\text{best} + c$}
  \STATE {\bf return} single hit at location with best score
\ELSIF{$d_\text{best} \le d_\text{max}$ or all seeds had $> h_\text{max}$ entries}
  \STATE {\bf return} multiple hits
\ELSE
  \STATE {\bf return} not found
\ENDIF
\end{algorithmic}
\end{algorithm}

This choice has two benefits. First, in many cases, the first candidate location
we find for a read is a good alignment, so we will compute the edit distance
to it faster than in $O(n^2)$ time. Indeed, 80\% of the time, the
first location we score for a read is the best, likely
because many of our seeds match only one location in the genome.
Second, because we are only interested in the best and second-best alignments,
we can lower $d_\text{limit}$ as we go along to check further candidate locations faster.
In particular, SNAP takes a parameter $c$, the \emph{confidence threshold}, that represents
the minimum difference in edit distance scores between the best and second-best alignments
that will let the algorithm return the best alignment rather than saying that the read maps
ambiguously. That is, if a read's best alignment is at position $p_1$ with edit distance
$d_1$, and its second-best alignment is at position $p_2$ with edit distance $d_2$, then
SNAP will return position $p_1$ if $d_2 - d_1 \ge c$ or report the read as ambiguous
otherwise. Thus, if we have a best match at some edit distance $d_\text{best}$, we can limit our
search to matches within distance $d_\text{best}+c-1$, as any poorer matches will
not affect our result. Furthermore, if we find a second-best match within this distance,
then we can set $d_\text{limit}$ to $d_\text{best}-1$, as our best hit so far cannot
be reported confidently and the only thing that would change our result is that
there is an undiscovered alignment with distance at most $d_\text{best}-c$ and no alignments with
distance within $c$ of that. 
With these optimizations, more than 90\% of SNAP's edit distance calls after the first reach
$d_\text{limit}$ without fully computing the distance and return early.

Finally, SNAP also counts the number of seeds that matched at each database location
to eliminate some candidates before scoring their edit distances. In particular,
if $t$ non-overlapping seeds from the read did not match a given location in the database,
then the reference sequence at that location must have edit distance at least $t$ from
the read~\cite{baeza-yates-seeds}. We use this observation in two ways.
First, after we have tested $t$  non-overlapping seeds, we know that the edit distance
to any location we have not yet found as a candidate is at least $t$, so if our best
alignment has distance $d_\text{best}\le t-c$, we need not try other seeds and can score all our
candidates instead. Second, as SNAP finds candidates, it only scores one after each
seed it tests (the one matching the most seeds), so it can eliminate some candidates
without ever scoring them.
We have also tried looking up more seeds before scoring the first candidate,
but surprisingly, this had little effect on performance. We are still investigating whether
it can be useful at some read lengths and error rates.

For simplicity, we have omitted a few other aspects of the algorithm from the pseudocode:
\begin{enumerate}
  \item When a read contains an insertion or deletion relative to the reference database, looking up different
  seeds from it may give slightly different alignment positions. For example, with a seed size of 20, perhaps
  the first seed of the read matches position $p$, while the second seed matches position $p+21$ because
  there was a deletion between the two seeds. In this case, it may appear that the read maps ambiguously,
  either at position $p$ or $p+1$. We avoid this by merging together results for positions that are close-by
  and only considering the best result.\footnote{
  Specifically, we group the positions in the genome into buckets of size 32 and only track the
  best aligning location within each bucket.}
  
  \item About half of the reads will be reverse complements relative to the reference database. To match
  these, we populate the index with both forward and reverse complement versions of each string.
  
  \item We choose the order of seeds from the read to minimize overlap between them. We start with as
  many sequential non-overlapping seeds as possible, then offset the
  next set of seeds by $\frac{s}{2}$, then $\frac{s}{4}$, $\frac{3s}{4}$, etc.
\end{enumerate}


\section{Ongoing Work}

After implementing the basic SNAP algorithm, we explored how various properties of its search parameters and input data affect its speed and accuracy.
In this section, we discuss our ongoing work to further improve SNAP based on this analysis.
We first outline observations about characteristics of the human genome that inhibit faster and more accurate alignment, creating a tradeoff between accuracy and speed.
Then, we describe our efforts to leverage knowledge about the genome to retain high accuracy without sacrificing speed.

\subsection{Observations}

\begin{figure*}
\centering
\includegraphics[scale=0.52]{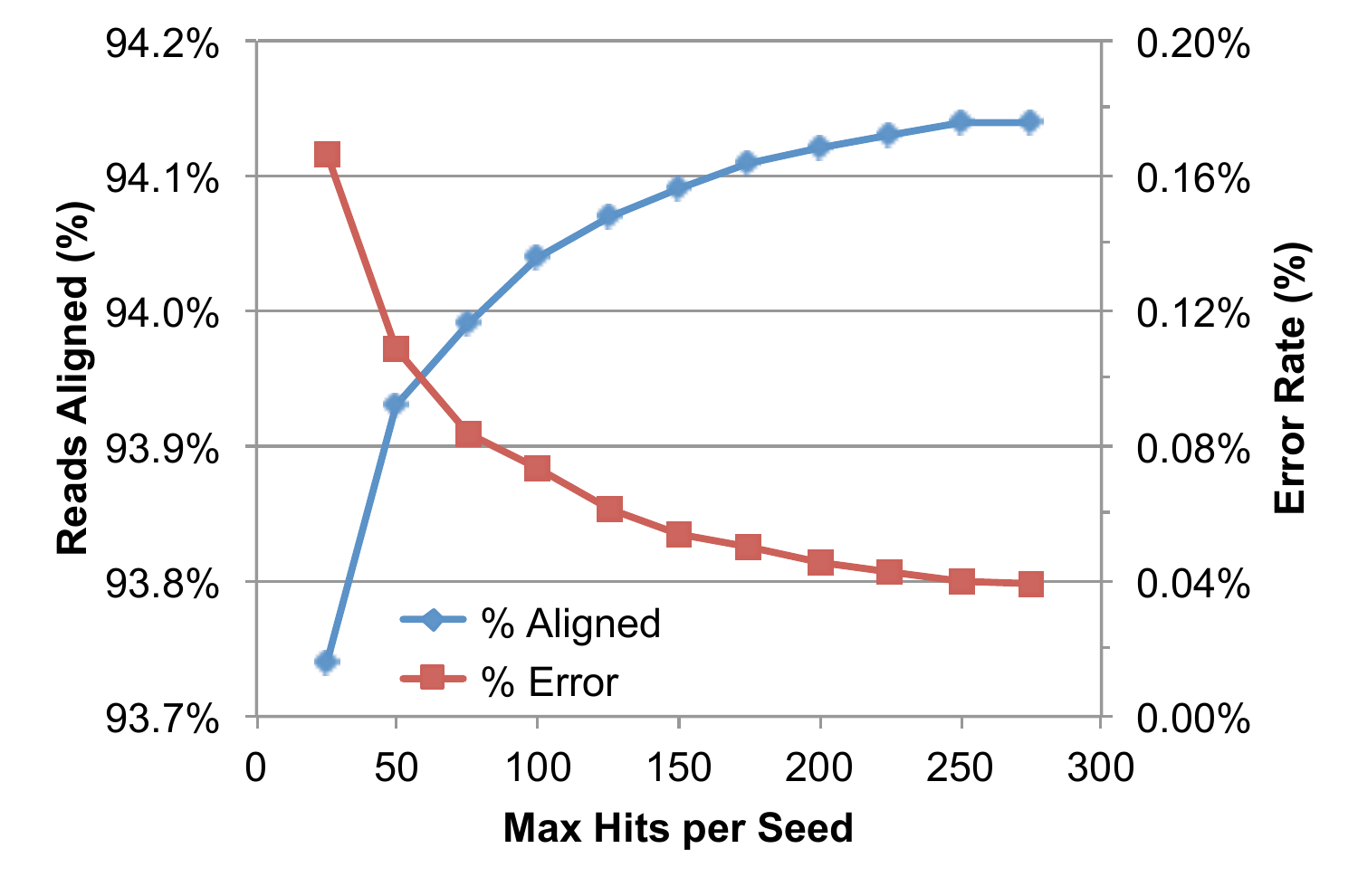}
\hspace{0.05in}
\includegraphics[scale=0.52]{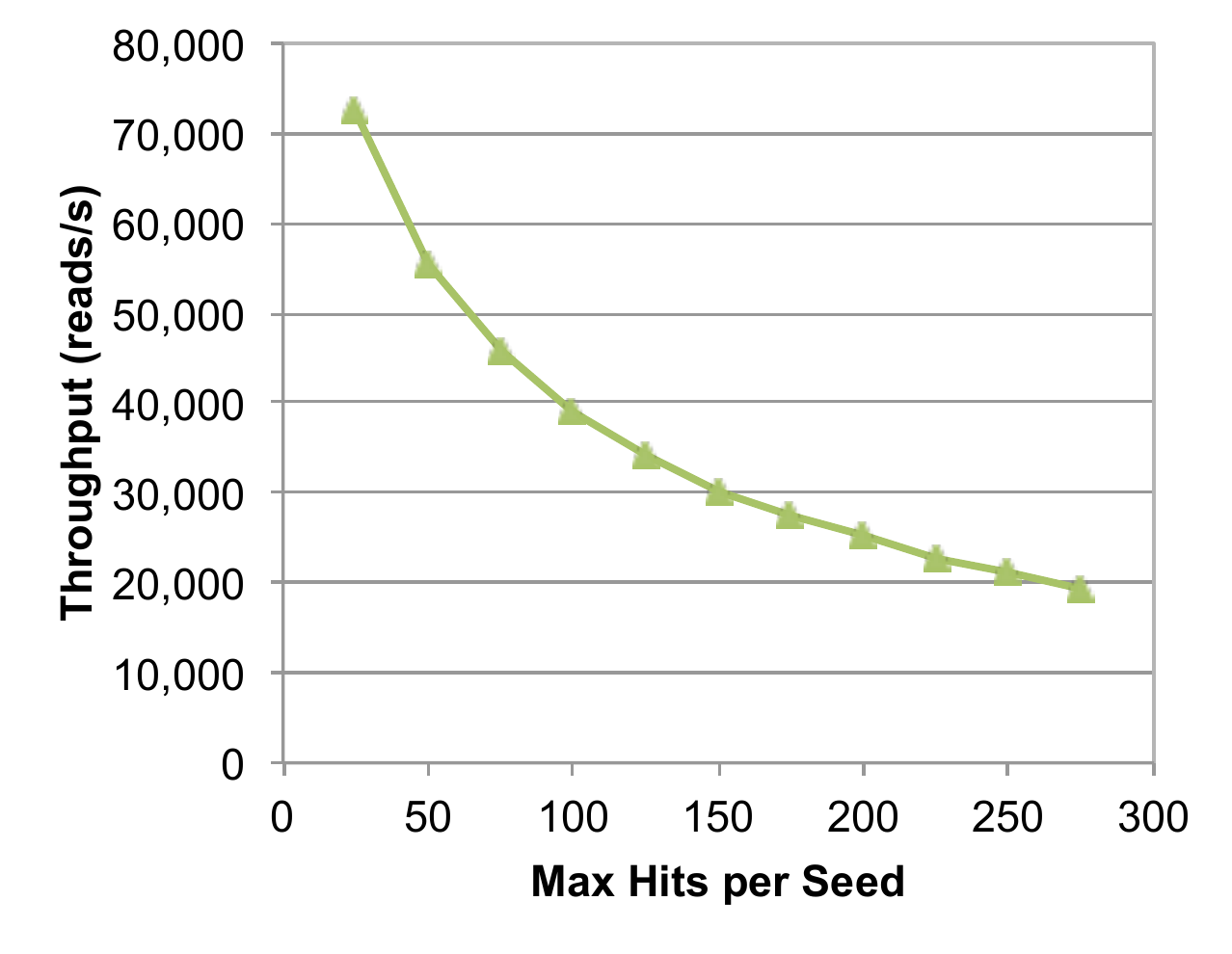}
\caption{Accuracy grows as we increase the number of locations we can check per seed, $h_\text{max}$, but throughput falls. \label{fig:hmax}}
\end{figure*}

To identify what was resulting in the most computational expense in our algorithm, we did a sweep over the algorithm parameters listed in Table \ref{tbl:parameters}.
We noticed that the only parameter that greatly affected our metrics of interest (reads aligned, error, and speed) was $h_\text{max}$, the maximum number of possible genome locations that we were willing to consider for each seed.
That is, exploring candidate locations from seeds with more hits yielded significantly more alignments and fewer errors, but at a cost of throughput.
Given that SNAP tries multiple seeds per read, this means that the reads in question mapped well to multiple locations. 
In other words, to align the last few percent of reads, we must test them against tens or hundreds of candidate
locations.
We plot this effect for 125 bp reads with a 2\% error rate in Figure~\ref{fig:hmax}.

Upon identifying this tradeoff between accuracy and throughput, we decided to learn more about how reads can match many  locations.
It is well known that the human genome is characterized by a great deal of redundancy.
Some of this is exact duplication, often in highly repetitive sequences.
For example, in chromosome 1 alone, the string {\tt AAAAAAAAAAAAAAAAAAAA} occurs over 36,000 times.
However, if a read matches multiple locations \emph{equally} well, it will quickly be discarded by SNAP
as ambiguous, so it should not affect the rate of confident alignments.

There has been less discussion about the fact that there is also substantial \emph{near-duplication} in the genome.
It is possible to find large groups of substrings where there are few exact duplicates, but all the strings differ from each other by very little.
An interesting feature of near duplicate regions is that there is not always a repetitive pattern.
Consider an example from chromosome 22.
We found one cluster containing over four hundred similar strings of length 100.
The ``consensus" string for the cluster, obtained by taking the most popular base at each position in the strings,
was {\tt
GCAAGCTCCGCCTCCCGGGTTCACGCCATTCTCC\linebreak
TGCCTCAGCCTCCCGAGTAGCTGGGACTACAGGCGCCC\linebreak
GCCACCACGCCCGGCTAATTTTTTGTAT}.
This string clearly has no repetitive pattern, but we found that
on average, the strings in the cluster differed from this consensus string by only six bases.\footnote{
Interestingly, \emph{which} of the six bases differ varies across strings, and for almost
all of the 100 possible positions, we found three or four distinct bases present across the set of strings.
}

The simplest way to avoid spending time on reads that match such clusters of similar regions is to impose a cutoff.
This is the idea behind restricting $h_\text{max}$, and it works well for exact duplicates, where all the
seeds have hundreds of hits.
However, this approach poses two problems with near-duplicates: we might set the threshold too low, and miss
some reads that actually align unambiguously to one location, or miss some locations where a read aligns and
return a confident result for it when it is in fact ambiguous; or we might set the threshold too high, and
incur a high computational cost checking reads against many locations.
This leads directly to the tradeoff between accuracy and speed observed in setting $h_\text{max}$.
To get around this constraint, our current work aims to identify the similar regions in the
reference database \emph{in advance} and treat matching against them specially.

\subsection{Approach}

To handle similar regions more effectively, we are pursuing an approach consisting of two steps:
(1) finding the similar regions in the genome and (2) matching against a cluster of such regions
efficiently.

Finding regions of exact duplication is simple; we just look for identical substrings of length \(r\), where \(r\) is the length of our reads.  This can be done by simply hashing these substrings. We can then save time in alignment by comparing against each repeated string only once.

Finding regions of near-duplication is more complicated, but parallels the approach taken for exact duplicates.
We still want to group substrings of length \(r\) via hashing.
However, since we are looking for near duplicates rather than exact duplicates, we do not hash on the entire substring.
Instead, we hash on \(C\) random columns (positions in the string).
Under this hashing scheme, similar substrings are more likely to hash together than random substrings.
However, the particular set columns chosen will influence which pairs of similar substrings hash together.
Therefore, we repeat the hashing many times, each time choosing an independent set of \(C\) columns
to hash on.
Then, any substrings that hash together significantly more often than random strings across these tests
will be considered similar.

After finding the clusters of similar regions, how can we use them during alignment?
We are exploring several approaches, but we believe that even simple techniques will help.
First, simply having the elements of a cluster together will reduce the error rate at low values of $h_\text{max}$ by
allowing us to check a read against all of the strings in a cluster; that is, we will no longer miss some good
alignments because they happened not to be found by our seeds, and mistakenly report a read as confident.
Second, putting the strings in the cluster in a contiguous location in memory will significantly reduce
CPU cache misses during alignment, which will save time as each cache miss costs the same amount of
time as hundreds of CPU cycles.
Finally, we have developed several algorithms that can match a read against a cluster of similar regions
faster than matching against each string individually, by exploring the inherent \emph{redundancy} between
the strings in the cluster. Intuitively, the mere fact that these strings are similar means that a significant
amount of computation can be shared in aligning against them.


\section{Related Work}

Since the development of shotgun sequencing methods, sequence alignment has received
considerable research interest. We discuss the major approaches taken to alignment as
well as several recent algorithms, but we refer the reader to~\cite{li-survey} for a
more comprehensive survey.

Most current aligners use one of two approaches: hash-based seed-and-extend methods
or prefix trie methods based on the Burrows-Wheeler transform (BWT).
The seed-and-extend method was pioneered by BLAST~\cite{blast}, which builds a hash
index of small substrings (seeds) of the reference genome and checks seeds from the
reads against it for exact matches.\footnote{Alternatively, one can build seeds from
the reads and check the genome against them.} It then uses a local extension
process at each seed hit to find good alignments. While BLAST and various further tools
refining its approach have been highly successful, they were found too slow
for the short read data generated by modern DNA sequencers, so they have generally been
replaced for these workloads by BWT-based aligners.

BWT-based aligners, including Bowtie, SOAP2, BWA, and BWA-SW~\cite{bowtie, soap2, bwa, bwa-sw},
align reads against a prefix trie of the reference genome, which is represented compactly
using the Burrows-Wheeler transform~\cite{bwt}.
While a prefix trie allows for fast exact string matching against a reference text, the
main challenge these algorithms address is inexact matching.
Botwie, SOAP2, BWA perform inexact matching through backtracking --- that is, by trying
to insert errors at various positions in the read as it is matched against the prefix
trie --- and cut the potentially exponential cost of the search through various
heuristics, such as limiting the number and types of errors considered.
This approach therefore comes at a cost of accuracy: certain types of errors and mutations
cannot be matched. For example, BWA assumes that there are no errors in the first 20 bases
of the read, while Bowtie does not allow indels.
In addition, backtracking scales poorly with longer reads: as the number of errors one
wishes to tolerate per read increases, the cost of backtracking grows exponentially.
BWA-SW addresses these problems for long reads by using Smith-Waterman alignment of
the suffix tree against a directed acyclic word graph (DAWG) of the read.
However, we found that SNAP performs substantially better than
BWA-SW and BWA for both short and long reads, while also providing higher accuracy.

SNAP is based on the seed-and-extend method, but it gains its speed from several
observations and optimizations. First, SNAP uses longer seeds than previous hash-based
aligners, which is feasible with today's sequencing technologies because reads are
getting longer. When algorithms like BLAST were developed, short reads were 20 to 30
base pairs in length, so seeds necessarily had to be short (\eg 10 bp), leading to
extraneous work to process false positive hits. Today, short reads are at least 100 bp
in length, and are expected to increase. This allows us to use longer seeds, which reduce
the number of alignments performed, as explained in Section~\ref{sec:methods-index}.
Second, SNAP uses an edit distance algorithm that runs faster when the strings match
closely~\cite{ukkonen}, because with long seeds, most reads will match well at any locations
where a seed hits. This substantially reduces the cost of local alignment. Finally,
SNAP takes advantage of the larger memory capacities available today to reduce the number
of hashtable lookups for its index. The result is an algorithm that runs faster than
existing hash and BWT based aligners without trading away accuracy. SNAP will find
alignments with arbitrary numbers and types of errors as long as at least one contiguous
seed in the read matches the reference genome, which happens for long enough reads even
at relatively high error rates.

Finally, another recently developed aligner that takes advantage of high memory is
WHAM~\cite{wham}. WHAM also uses a hash-based index and eliminates some candidates using
the count of matching seeds, but it searches for multiple small seeds
from the read instead of a larger seed like SNAP's. The WHAM algorithm tolerates
fewer errors per read than SNAP --- only up to 5 errors in total and 3 gaps --- resulting
in lower accuracy. Furthermore, its performance decreases quickly with the number of
errors and gaps. We tested WHAM version 0.1.2 against a simulated dataset of 70 bp reads
with 2\% error from the human genome and found that it only aligned 40\% of reads
with its default setting of up to 3 errors, at a speed of 30,000 reads/s. Increasing the
number of errors and gaps to the maximum supported raised the number of reads aligned
to 60\% but reduced the speed to 4000 reads/s. In contrast, SNAP aligns 86.7\%
of this dataset with 0.08\% error at a speed of 52,000 reads/s.


\section{Conclusion}

As DNA sequencing technology continues to improve faster than Moore's law,
it is opening up new medical and scientific applications.
However, processing the short read data generated by this technology is a
growing computational challenge.
To address this challenge, we have presented SNAP, a new short and long read
aligner that runs 10--100$\times$ faster than state of the art tools while
providing higher accuracy.
SNAP derives its performance from a careful cost analysis of the seed-and-extend
alignment method and optimizations that greatly accelerate it for today's
longer read lengths and larger memory capacities.
We calculate that SNAP can align reads with 30$\times$ coverage of a
human genome in less than an hour on Amazon EC2, for a cost of \$2 .
Most importantly, our experience with SNAP shows that a careful
reconsideration of sequence processing algorithms in light of today's computer
hardware and sequencing technologies can yield substantial speedups.

\section{Acknowledgements}

We would like to thank Matthew Meyerson for providing motivation and encouragement for this work, and for his feedback on its utility in cancer and pathogen discovery.
This research is supported in part by gifts from the following Berkeley AMP Lab sponsors: Google, SAP, Amazon Web Services, Cloudera, Ericsson, General Electric, Huawei, IBM, Intel, Mark Logic, Microsoft, NEC Labs, Network Appliance, Oracle, Quanta Computer, Splunk and VMware, and by DARPA (contract \#FA8650-11-C-7136).

{\small
\raggedright
\bibliographystyle{abbrv}
\bibliography{bibinfo}
}

\end{document}